\documentclass{article}%
\usepackage{amsfonts}
\usepackage{amsmath}
\usepackage{amssymb}
\usepackage{graphicx}%
\begin{document}

\title{Phase Space Analysis of Fluorine-Oxygen-Nitrogen Network and Energy Generation
in $^{18}F(p,\alpha)^{15}O$ Reaction}
\author{Babur M. Mirza\\Department of Mathematics, \\Quaid-i-Azam\ University, {45320 Islamabad}, Pakistan\\E-mail: bmmirza2002@yahoo.com}
\maketitle

\begin{abstract}
Reaction network of fluorine-$18$, oxygen-$15$ and nitrogen-$15$ is considered
for its temperature dependent energy output. The main reactions for generation
and annihilation of oxygen and fluorine are coupled in the reaction equations
while nitrogen is produced as a decay product. We find that the governing set
of equations for $^{18}F(p,\alpha)^{15}O$ process in the phase diagram exhibit
a predominance of the direct reaction $^{18}F+p\rightarrow^{15}O+\alpha$
rather than the reverse reaction consuming $^{15}O$. The time-scale determined
by the exact solution of this system yields relatively short time-scale
conversion of fluorine into oxygen indicating an energy generation of $2.8MeV$
per reaction. The temperature dependence shows that the effective reaction
occurs at temperature about $0.04GK$ or above.

\end{abstract}

\section{Introduction}

Energy generation by nuclear processes involves nuclear transmutations of
elements depending upon their respective reaction rates. The production and
destruction of elements and the time-evolution of resulting nuclear abundances
is modelled by reaction networks that include both sink and source terms. In
general, an exothermic process can be defined as efficient if the reaction
occurs at low temperature and short time-scale. To determine efficiency and
the time-evolution of each nuclear reaction in a network the general system of
differential equations for nuclear abundances is solved. Such a system is
generally coupled and is highly sensitive to initial conditions (Hix \&
Thielemann 1999; Martin et. al. 2018; Boehnlein et.al. 2022) for each
abundance. Various improved computational techniques (see for example,
Longland et. al. 2014; Arbey et.al. 2020; and references therein) are used to
solve the general stiff system of equations. However, it is generally known
that the results are extremely sensitive to computational error, and can
become unstable in more complex cases (M\"{u}ller 1986; Calder et. al. 2002).
It is therefore difficult to trace the details of each process involved in general.

For relatively simpler cases, mainly in astrophysical environments, the
network equations successfully model the production of lighter elements
(Clayton 1983; Hix \& Meyer 2006; Lippuner \& Roberts 2017; Seitenzahl \&
Townsley 2017; Thielemann 2019; Iliadis \& Coc 2020; Bruenn et. al. 2020). The
temperatures required are in some cases ($p-p$ reaction, and the $CNO$ cycle)
comparable to $10^{6}K$ or above, however at these temperatures sufficient
amount of hydrogen can be fused over very long time-scales ($\sim10^{7}yr$ for
hydrogen burning and for the CNO cycle at a time scale of $\sim10^{10}yr$).
For heavier nuclei, such as oxygen and silicon and elemants above silicon, the
time-scales are relatively shorter, approximately $6$ months for oxygen and
$1$ day for silicon, however the temperature required for ignition are in
excess of $10^{9}K$ (in main sequence stars). Beyond the $Fe$-peak large
networks are generally solved using various approximations to reduce the
computational procedure.

Apart from the problem of origin of elements, the reaction networks control
the production and destruction of elements in a nuclear process. In such cases
each transmutation occurs in a nuclear reaction $A+b\rightarrow c+D$;
abbreviated in the following as $A(b,c)D$, where $A$ is the target nucleus,
$b$ is the projectile, $c$ is the ejectile, and $D$ is the product nucleus. In
production of different elements the network equations couple the forward and
backward reactions as well as the neighboring reactions that act either as a
source or a sink. The abundance or concentration of an element in each step,
within a given time-scale, thus measures the effectiveness of a specific
nuclear reaction. As pointed out, it is not possible in general to identify
efficiency of each process due to the complexity of the coupled system of equations.

Global phase space methods allow the analysis of each network component
separately, hence enables the estimation of efficiency of each process within
the network. In this procedure only the neighboring interactions are retained
and analyzed in phase space for each pair of reactants. Here we identify
fluorine-oxygen reaction $^{18}F(p,\alpha)^{15}O$ as an efficient reaction in
the production of $^{15}O$ isotope over relatively short period of time. This
process is interesting for that in a nuclear network, not only the reverse
reaction $^{15}O+\alpha\rightarrow^{18}F+p$ is involved, but the $^{15}O$
isotope undergoes a $\beta$-decay also via $^{15}O\rightarrow^{15}N+e^{+}+\nu
$, producing $^{15}N$ in relatively short time (approximately in $122s$),
which might inhibit the production of the oxygen isotope. However, it is shown
in the following sections that at relatively low temperatures ($\lesssim
0.4GK$), the production of $^{15}O$ overtakes the production of $^{18}F$ after
a sufficient time of fuel consumption ($\gtrsim10^{3}s$), hence making the
forward reaction a possible efficient energy generation process.

\section{Mathematical Formulation}

The governing equations for network reactions are coupled via the source and
sink terms. If $N_{i}$ is the number density of the nuclei $i$, then the
abundance evolution is given by the system of coupled differential equations,%
\begin{align}
\frac{dN_{i}}{dt}  &  =\left[  \sum_{j,k}N_{j}N_{k}\left\langle \sigma
v\right\rangle _{jk\rightarrow i}+\sum_{l}\lambda_{\beta},_{l\rightarrow
i}N_{l}+\sum_{m}\lambda_{\gamma},_{m\rightarrow i}N_{m}\right]  -\nonumber\\
&  \left[  \sum_{n}N_{n}N_{i}\left\langle \sigma v\right\rangle _{ni}+\sum
_{o}\lambda_{\beta},_{i\rightarrow o}N_{i}+\sum_{p}\lambda_{\gamma
},_{i\rightarrow p}N_{i}\right]  , \tag{1}%
\end{align}
where the first and the second parenthesis represent all processes for the
production and destruction of the $i$th nuclei, respectively. The first
parenthesis includes the following possible types of sums: $\sum_{j,k}=$ sum
over all reactions producing nucleus $i$ via reaction between $j$ and $k$;
$\sum_{l}=$sum over all $\beta$-decays of nuclei $l$ leading to $i$; $\sum
_{m}=$sum over all photodisintegrations of nuclei $m$ leading to $i$. In the
second parenthesis the corresponding sums (destruction terms) have the
equivalence as: $\sum_{n}\leftrightarrow\sum_{j,k}$, $\sum_{o}\leftrightarrow
\sum_{l}$, and $\sum_{p}\leftrightarrow\sum_{m}$. In place of the number
densities $N_{i}$ in a nuclear reaction, it is advantageous to use the mole
fractions $Y_{i}=N_{i}/\rho N_{A}$, $N_{A}$ being the Avogadro number.

For the $^{18}F(p,\alpha)^{15}O$ reaction, the immediate reaction for the
destruction terms is the reverse reaction $^{15}O+\alpha\rightarrow$
$^{18}F+p$,\ \ and there is the decay$\ ^{15}O\rightarrow^{15}N+e^{+}+\nu$.
The resulting set of equations therefore is given by,\bigskip%
\begin{align}
\frac{d^{18}F}{dt}  &  =-H^{18}F\left\langle \sigma v\right\rangle
_{^{18}F(p,\alpha)}+^{4}He^{15}O\left\langle \sigma v\right\rangle
_{^{15}O(\alpha,p)},\nonumber\\
\frac{d^{15}O}{dt}  &  =-^{15}O^{4}He\left\langle \sigma v\right\rangle
_{^{15}O(\alpha,p)}+^{18}FH\left\langle \sigma v\right\rangle _{^{18}%
F(p,\alpha)}-\lambda_{\beta,^{15}N}{}^{15}O,\nonumber\\
\frac{d^{15}N}{dt}  &  =-^{15}NH\left\langle \sigma v\right\rangle
_{^{15}N(p,\alpha)}+^{12}C^{4}He\left\langle \sigma v\right\rangle
_{^{12}C(\alpha,p)}. \tag{2}%
\end{align}
Here the last equation is decoupled, and the remaining system of equations has
the matrix form,%
\begin{equation}
\mathbf{\dot{X}}=A\mathbf{X}, \tag{3a}%
\end{equation}
where $\mathbf{X}=\left[
\begin{array}
[c]{cc}%
^{18}F & ^{15}O
\end{array}
\right]  ^{T}$, $\mathbf{\dot{X}}=d\mathbf{X}/dt$ and $A$ is the coefficient
matrix, given by
\begin{equation}
A=\left[
\begin{array}
[c]{cc}%
-H\left\langle \sigma v\right\rangle _{^{18}F(p,\alpha)} & ^{4}He\left\langle
\sigma v\right\rangle _{^{15}O(\alpha,p)}\\
H\left\langle \sigma v\right\rangle _{^{18}F(p,\alpha)} & -^{4}He\left\langle
\sigma v\right\rangle _{^{15}O(\alpha,p)}-\lambda_{\beta,^{15}N}%
\end{array}
\right]  . \tag{3b}%
\end{equation}
The governing system of equations has the phase space variables $^{18}%
F-^{15}O$, which we call the fluorine-oxygen system.

\section{Phase Space Analysis}

Phase space method allows the dividing of solutions of equations of type (3)
according to their time-evolution in phase space. Since this can be done
independently of the initial conditions involved the analysis presents a
robust estimate of the resulting trend at any given temperature. Solutions are
studied depending on their respective eigenvalue-eigenvector pairs
respectively (Davis 1962; Mirza 2015).

The characteristic equation $\det\left\vert A-\lambda I\right\vert $ for the
fluorine-oxygen system (3) has the eigenvalues,%
\begin{equation}
\lambda_{1},_{2}=\frac{1}{2}((a+c)\pm\sqrt{(a+c)^{2}-4(ac-bd)}), \tag{4}%
\end{equation}
where $a=-H\left\langle \sigma v\right\rangle _{^{18}F(p,\alpha)}=-d$,
$b=^{4}He\left\langle \sigma v\right\rangle _{^{15}O(\alpha,p)}$,
$c=-b-\lambda_{\beta,^{15}N}$. The general solution to the system (3) is then
given by%
\begin{equation}
\mathbf{X}(t)=\mathbf{C}_{1}e^{\lambda_{1}t}+\mathbf{C}_{2}e^{\lambda_{2}t}.
\tag{5}%
\end{equation}
The system has the only critical point (node) at $(^{18}F_{c},^{15}%
O_{c})=(0,0)$. In general the eigenvalues can be expressed as $\lambda
_{1},_{2}=(\mu\pm i\omega)/2$, where $\mu=(a+c)$ and $\omega=\sqrt
{4(ac-bd)-(a+c)^{2}}$ are both real in cases 1 to 4 below. The following cases
may arise as time $\ t$ increases.

\textit{Case 1:} $\lambda_{1}>\lambda_{2}>0$ gives an unbounded solution, so
that the node is an improper critical point. The system is unstable, as the
length of phase space vector $\mathbf{X}(t)$ increases without bound as
$t\rightarrow\infty$. All orbits move away from the origin. In this case the
eigenvalues are real, therefore $4(ac-bd)<(a+c)^{2}$. This gives the condition that%

\begin{equation}
-4(a\lambda_{\beta,^{15}N})>(a-b-\lambda_{\beta,^{15}N})^{2}. \tag{6}%
\end{equation}
Both eigenvalues are positive, which implies that $\mu\pm\omega>0$. Since both
$a$ and $\ c$ are negative, $\mu<0$. Therefore unbounded solutions are
possible only when $\omega>\mu$ or equivalently $-bd>a^{2}+c^{2}$. As $b>0$
and $d>0$, this is not possible. Thus there are no unbounded solutions, and
production of elements terminates in time.

\textit{Case 2: } $\lambda_{1}<\lambda_{2}<0$ gives a bounded solution. The
node is a proper critical point. Trajectories move towards the origin as
$t\rightarrow\infty$. The system is stable and the length of phase space
vector $\mathbf{X}(t)$ decreases with time. The eigenvalues are real,
therefore $4(ac-bd)<(a+c)^{2}$ yields the same condition as above. However,
$\mu\pm\omega<0$ implies no contradiction, therefore bounded solutions are possible.

\textit{Case 3: }If $\lambda_{1}<0$ and $\lambda_{2}>0$ (or vice-versa) the
solution is unstable. The node is a saddle point. As $t\rightarrow\infty$, the
orbits move towards the origin along the $X_{1}$ eigenvector and away from the
origin along the $X_{2}$ eigenvector (or conversely $\lambda_{1}<0$ and
$\lambda_{2}>0$ for $\lambda_{1}>0$ and $\lambda_{2}<0$ ). The eigenvalues are
real, therefore $4(ac-bd)<(a+c)$. As before these solutions are possible,
however unstable.

\textit{Case 4: }For the degenerate case $\lambda_{1}=\lambda_{2}$, $\omega
=0$, and the condition $4(ac-bd)=(a+c)^{2}$ is obtained. This gives,
\begin{equation}
-4(a\lambda_{\beta,^{15}N})=(a-b-\lambda_{\beta,^{15}N})^{2}. \tag{7}%
\end{equation}
Further, $\lambda_{1}=\lambda_{2}=\mu=-(a+c)/2$. Since there is only one
eigenvalue-eigenvector pair in this case, degenerate eigenvalues can be
discarded. Therefore there are no periodic solutions for the fluorine-oxygen system.

The following cases correspond to complex eigenvalues. Thus $\omega$ is real,
and $-4(a\lambda_{\beta,^{15}N})<(a-b-\lambda_{\beta,^{15}N})^{2}$ for cases 5
to 7.

\textit{Case 5:} If $\mu=0$, the solution is periodic, with period $2\pi
\omega$. This corresponds to the condition that $a=-c$. Periodic solutions are
excluded since both $a$ and $c$ are negative above.

\textit{Case 6:} If $\mu<0$, the orbit is an inward spiral, and the origin is
a stable focus. Equivalently this means that $a<-c$. Since $a<0$ and $c<0$,
solution can be an inward spiral.

\textit{Case 7:} If $\mu>0$, the solution is an outward spiral, with origin as
unstable focus. The corresponding condition is that $a>-c$, which is not possible.

Thus we conclude that for system (3) only possible branches of solution can be
case 2, 3 and 6 only; namely, a bounded solution with node as a critical
point, unstable solution with node as a saddle point, and an inward spiral
with origin as unstable focus. This gives only one stable solution for the
case (2) above.

In the following we investigate the solutions of system (3) using input data
for the reaction rates. In each reaction number density of hydrogen and helium
nuclei is normalized, therefore we have $H$ and $He$ equal to unity. The
reaction rates are calculated using STARLIB online data-base (Sallaska et. al.
2013) for forward, backward and decay reactions.

\section{Results}

Figure 1 show the phase space plot between flourine-$18$ and oxygen-$15$
abundance at $0.1GK$ temperature. In this case $a=-8.4733\times10^{-5}%
cm^{3}/mol\cdot s$, and $c=-5.68155\times10^{-3}cm^{3}/mol\cdot s$. The
eigenvalues ($-0.005681,-0.000084$) are both negative, and the node is a
proper critical point. This is case (2) above, hence a stable solution is
obtained independent of the initial conditions. In each of the following
cases, $b$ is practically zero for temperatures less than $1GK$.

The plot shows that initially there is a rapid increase in the fluorine
production and almost no increase or decrease in the production of the oxygen
isotope. The constant rate at which $^{15}O$ is maintained suggests that the
direct process $^{18}F+p\rightarrow$ $^{15}O+\alpha$\ \ and the decay$\ ^{15}%
O\rightarrow^{15}N+e^{+}+\nu$ tend to maintain the production of oxygen in the
initial process. Later on the fluorine nuclei are consumed at a much faster
rate, indicating that the forward process becomes more effective.\ \ \ To
further understand the mechanism of oxygen production we plot the abundance as
functions of time (in seconds) in Figure 2 and 3. In Figure 2 we see that the
oxygen content is actually reduced dramatically as the process of fusion
starts. It then stabilizes and decreases at a slower rate afterwards.
Comparing with the consumption of fluorine (Figure 3), there is a gradual
decrease in fluorine content over time. However, it can be noticed that oxygen
burning in the same time interval after the initial rapid decrease is much
slower than fluorine burning in the same duration. This indicates that the
forward reaction $^{18}F+p\rightarrow$ $^{15}O+\alpha$ overtakes the reverse
reaction as well as the decay of oxygen into $^{15}N$ in due time.

This trend is also observable at higher temperatures, as shown in Figures 4
and 5. At $1GK$ temperature, fluorine reduction is about $10^{7}$ times faster
than the consumption of oxygen.

At lower temperatures (Figure 6 and 7), however, the trend is the same as seen
above. Oxygen production shows a sharp decrease at first, and then is consumed
at a slower rate, although production is decreasing still. The decrease is
comparatively small, so that oxygen is consumed more slowly than fluorine per second.

\section{Conclusions}

The faster consumption of fluorine implies that the forward reaction
$^{18}F+p\rightarrow$ $^{15}O+\alpha$ is eventually favoured in the reaction
network. The initial sharp decrease in the oxygen isotopes in the reverse
reaction at lower temperatures is due to the $\beta$-decay of $^{15}O$ into
$^{15}N$. This trend is suppressed at high temperatures. Even at lower
temperatures the $^{18}F+p\rightarrow$ $^{15}O+\alpha$ reaction is far more
efficient than the reverse reactions that reduce the production of $^{15}O$.
However, this needs longer time period for fuel to burn. After the initial
decrease in the fluorine fuel, the forward reaction can produce sustained
fusion for sufficient amount of time at temperatures less than that for the
common D-T or D-D reactions for which temperature higher than $0.8GK$ and high
pressure/density conditions are required. Extrapolating the above analysis,
the indicated trend in the $^{18}F(p,\alpha)^{15}O$ process, is expected to
yield $p+^{18}F$ fusion as an efficient process even at temperatures lower
than $0.04GK$. Provided longer confinement of the input nuclei, and sufficient
fluorine fuel to exceed the initial rapid burning, an energy output $Q=2.8MeV$
per forward reaction can be obtained in the form of $\alpha$ particle production.

\bigskip

FIGURE CAPTIONS:

Figure 1: Phase space diagram for the fluorine-oxygen system at temperature
$T=0.1GK$,. The abundance is scaled by the proton abundance for both elements.
The parameters in the governing set of equations are $a=-8.4733\times
10^{-5}cm^{3}/mol\cdot s$, $c=-5.68155\times10^{-3}cm^{3}/mol\cdot s$.

Figure 2: Oxygen production rate scaled by the hydrogen abundance at
temperature $T=0.1GK$. Parameters used are the same as in Figure1 above.

Figure 3: Corresponding fluorine production rate (scaled) at temperature
$T=0.1GK$. Parameters used are the same as in Figure1 above.

Figure 4: Oxygen production rate for temperature $T=1GK$, and
$a=-47637.0595cm^{3}/mol\cdot s$, $b=7.0041\times10^{-11}cm^{3}/mol\cdot s$,
$c=-5.68155\times10^{-3}cm^{3}/mol\cdot s$.

Figure 5: Corresponding to parameters in Figure 3 above, the production rate
for fluorine is exhibited here.

Figure 6: Production rate of $^{15}O$ for lower temperature $T=0.04GK$, and
the reaction rate parameters $a=-3.3466\times10^{-9}cm^{3}/mol\cdot s$,
$c=-0.00568cm^{3}/mol\cdot s$.

Figure 7: Fluorine production rate per unit time at temperature $T=0.04K$.
Parameters as in Figure 6 above.


\begin{thebibliography}{99}                                                                                               %


\bibitem {1}Arbey, A., Auffinger, J., Hickerson, K.P. and Jenssen, E.S., 2020.
AlterBBN v2: A public code for calculating Big-Bang nucleosynthesis
constraints in alternative cosmologies. Computer Physics Communications, 248, 106982.

\bibitem {2}Boehnlein, A., et. al., 2022. Colloquium: Machine learning in
nuclear physics. Reviews of Modern Physics, 94, 031003.

\bibitem {3}Bruenn, S.W., et. al., 2020. CHIMERA: a massively parallel code
for core-collapse supernova simulations. The Astrophysical Journal Supplement
Series, 248, 11.

\bibitem {4}Calder, A.C., et. al., 2002. On validating an astrophysical
simulation code. The Astrophysical Journal Supplement Series, 143, 201.

\bibitem {5}Clayton, D.D., 1983. Principles of Stellar Evolution and
nucleosynthesis (University Press, Chicago).

\bibitem {6}Davis, H.T., 1962. Introduction to Nonlinear Differential and
Integral Equations (Dover Publications, New York).

\bibitem {7}Hix, W.R. and Thielemann, F.K., 1999. Computational methods for
nucleosynthesis and nuclear energy generation. Journal of Computational and
Applied Mathematics, 109, 321.

\bibitem {8}Hix, W.R. and Meyer, B.S., 2006. Thermonuclear kinetics in
astrophysics. Nuclear Physics A, 777, 188.

\bibitem {9}Iliadis, C. and Coc, A., 2020. Thermonuclear reaction rates and
primordial nucleosynthesis. The Astrophysical Journal, 901, 127.

\bibitem {10}Lippuner, J. and Roberts, L.F., 2017. SkyNet: a modular nuclear
reaction network library. The Astrophysical Journal Supplement Series, 233, 18.

\bibitem {11}Longland, R., Martin, D. and Jos\'{e}, J., 2014. Performance
improvements for nuclear reaction network integration. Astronomy \&
Astrophysics, 563, A67.

\bibitem {12}Martin, D., Jos\'{e}, J. and Longland, R., 2018. On the
parallelization of stellar evolution codes. Computational Astrophysics and
Cosmology, 5, 10.

\bibitem {13}Mirza, B.M., 2015. Formation of secondary critical points in
thermally conducting interstellar gases. New Astronomy, 35, 20.

\bibitem {14}M\"{u}ller, E., 1986. Nuclear-reaction networks and stellar
evolution codes-The coupling of composition changes and energy release in
explosive nuclear burning. Astronomy and Astrophysics, 162, 103.

\bibitem {15}Sallaska, A.L., Iliadis, C., Champange, A.E., Goriely, S.,
Starrfield, S. and Timmes, F.X., 2013. STARLIB: a next-generation
reaction-rate library for nuclear astrophysics. The Astrophysical Journal
Supplement Series, 207, 18. https://starlib.github.io/Rate-Library/

\bibitem {16}Seitenzahl, I.R. and Townsley, D.M., 2017. Nucleosynthesis in
thermonuclear supernovae. arXiv preprint arXiv:1704.00415.

\bibitem {17}Thielemann, F.K., 2019. Explosive Nucleosynthesis: What we
learned and what we still do not understand. in Nuclei in the Cosmos XV
(Springer, New York).
\end{thebibliography}
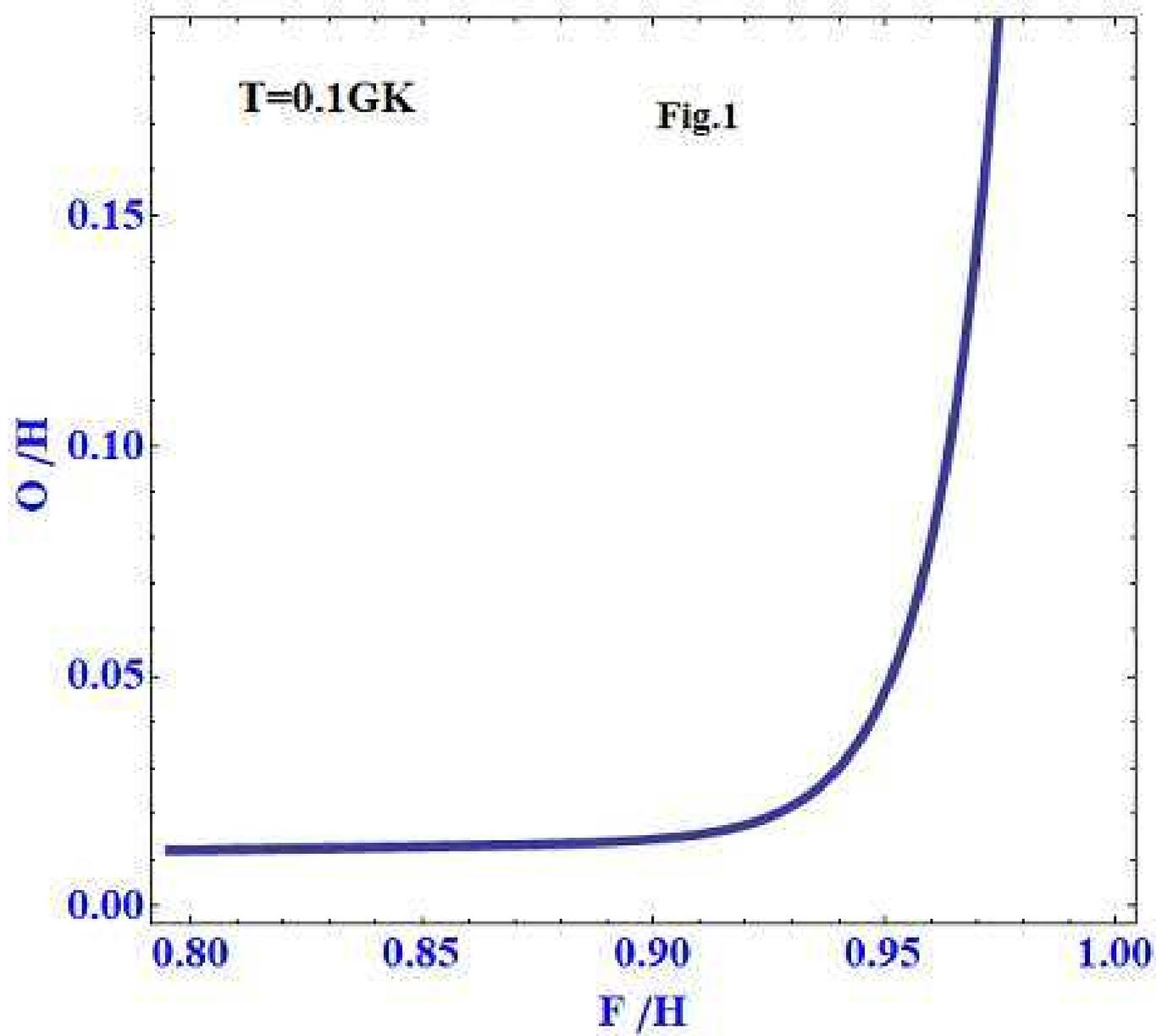
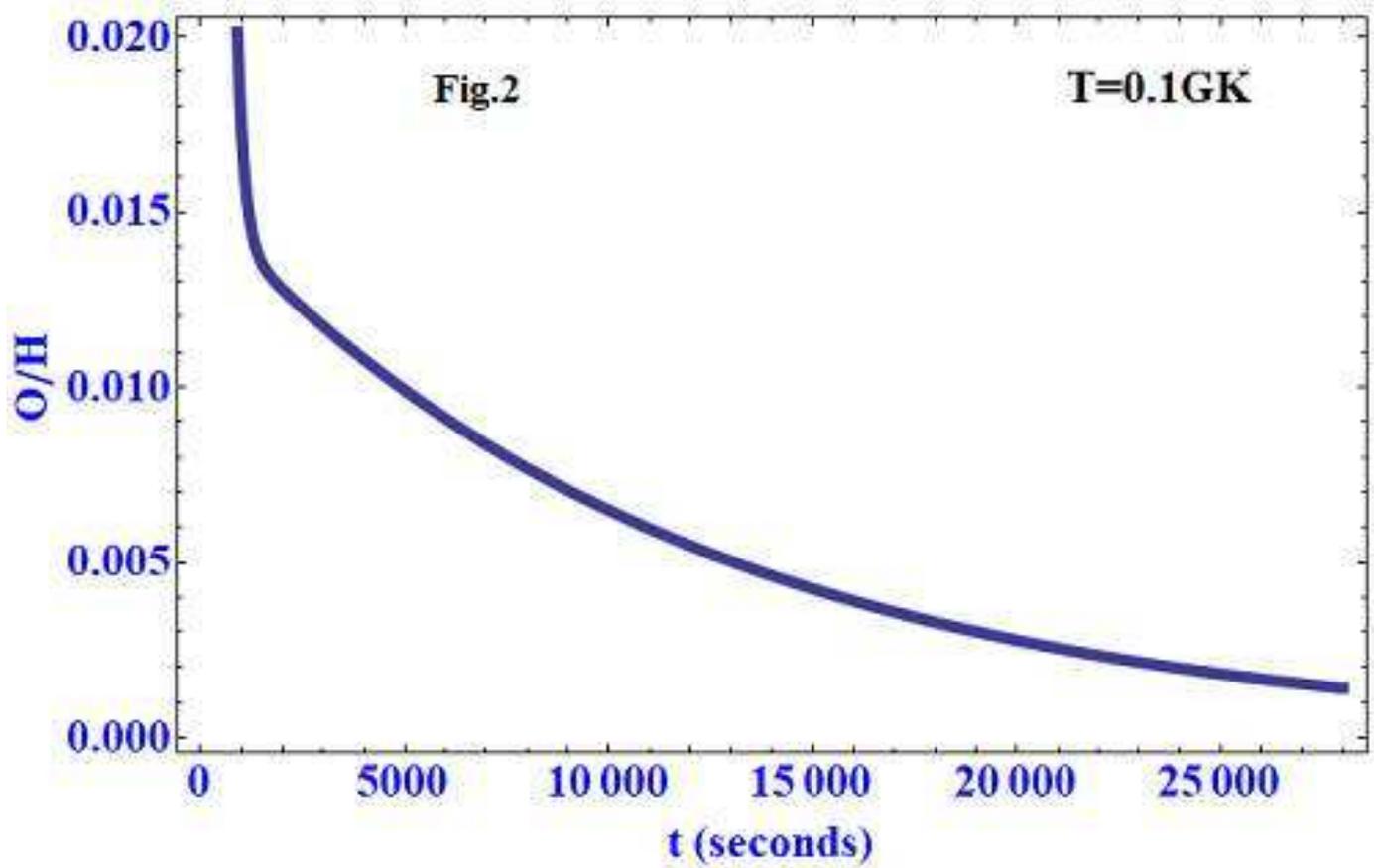
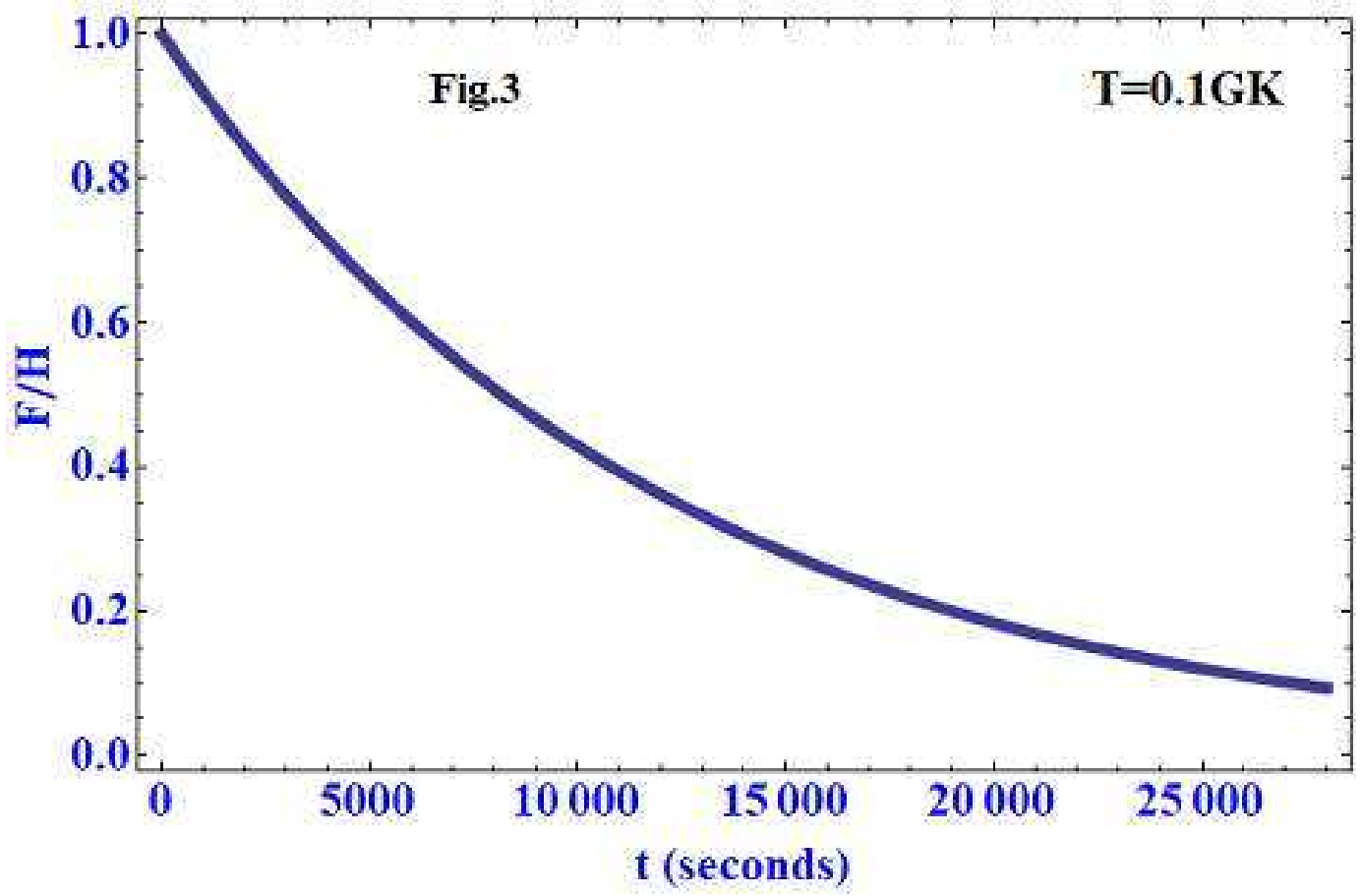
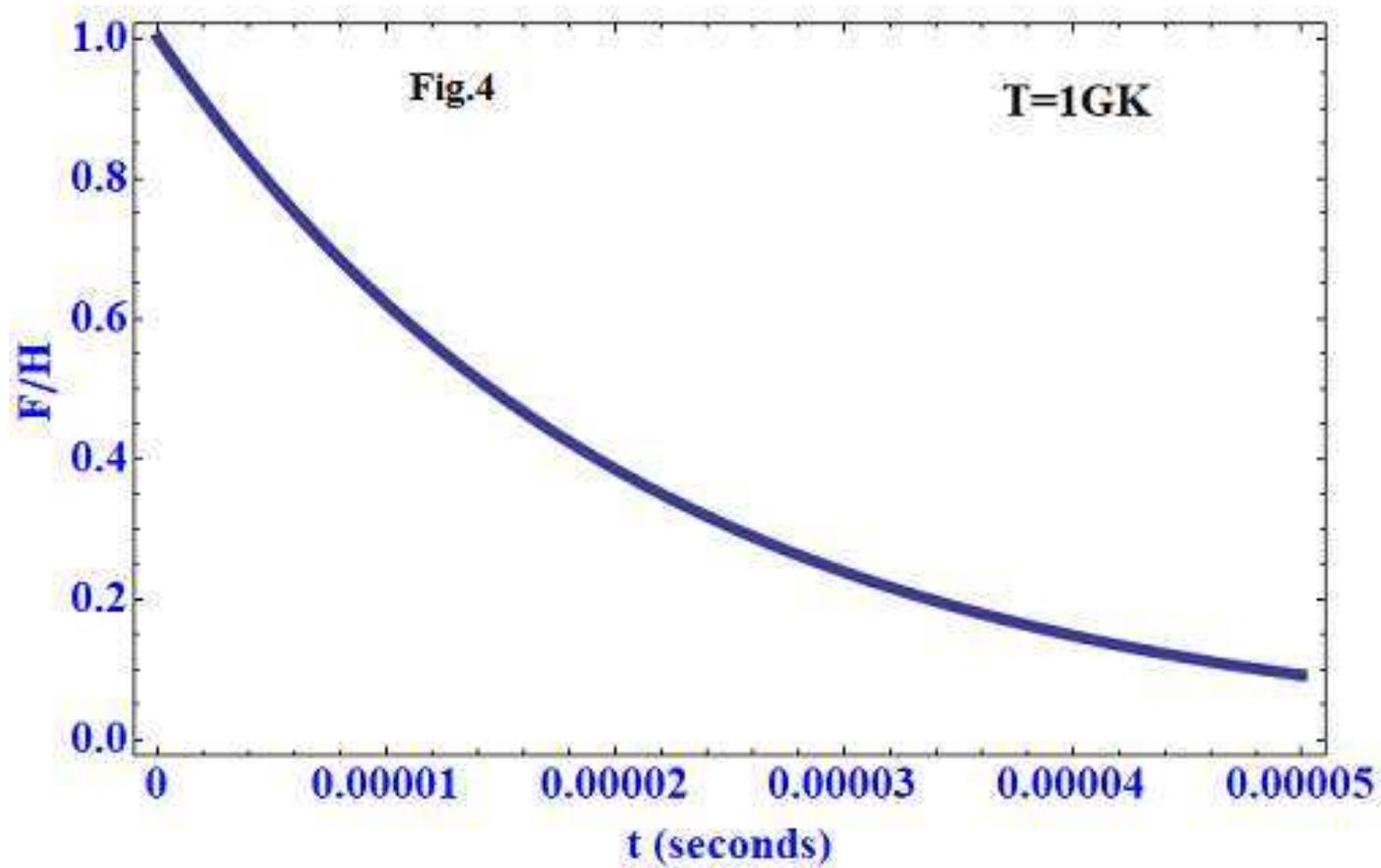
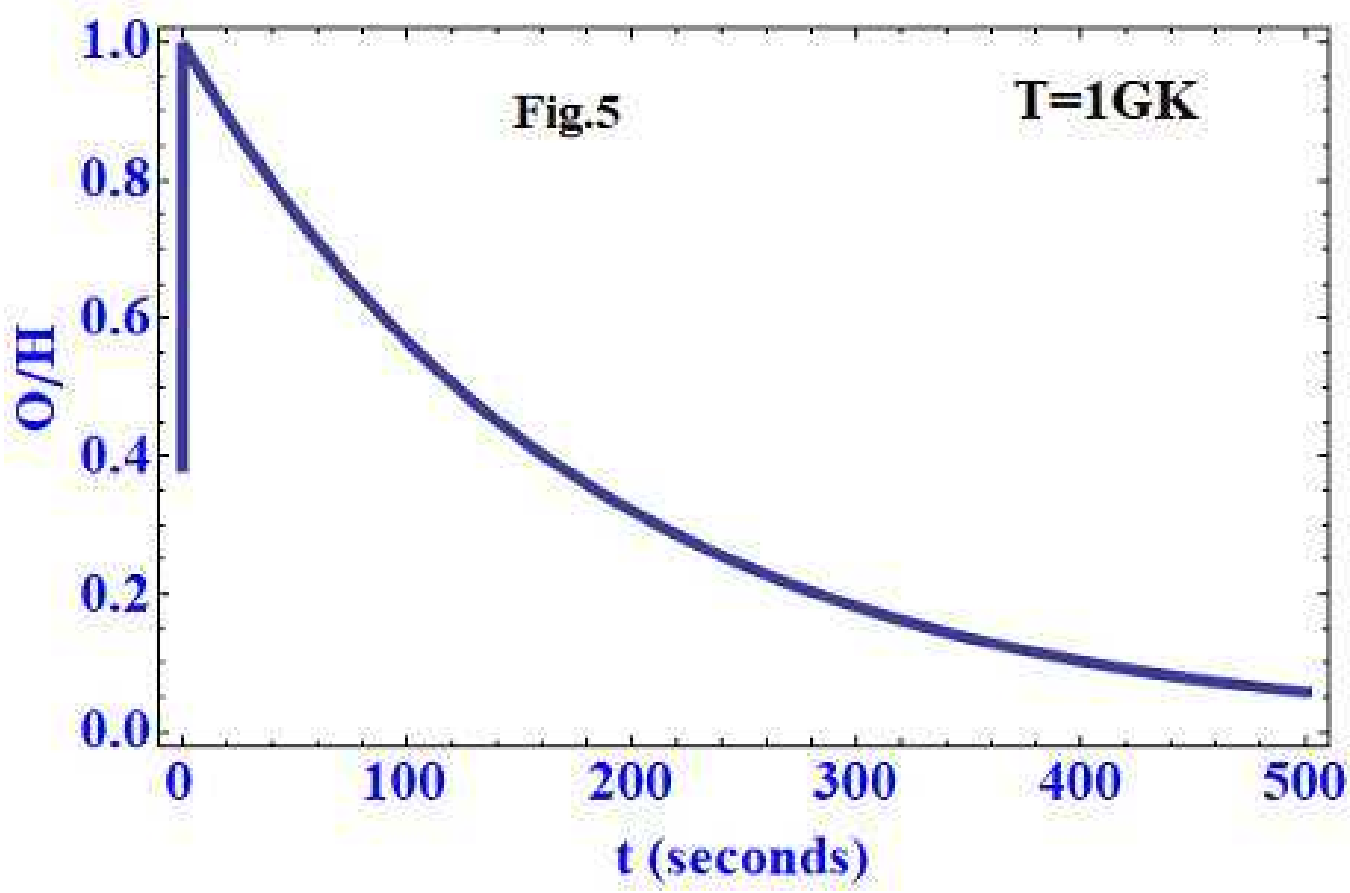
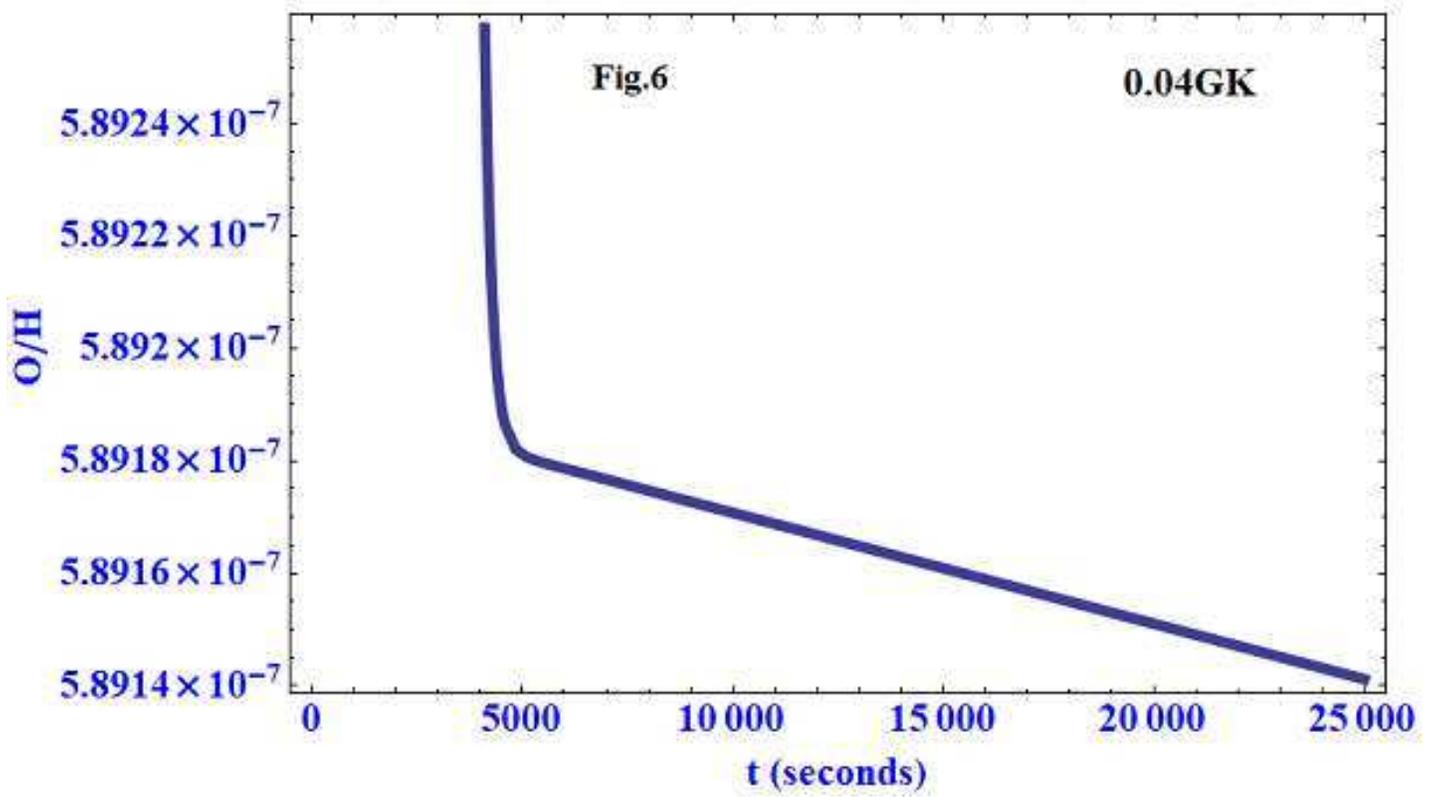
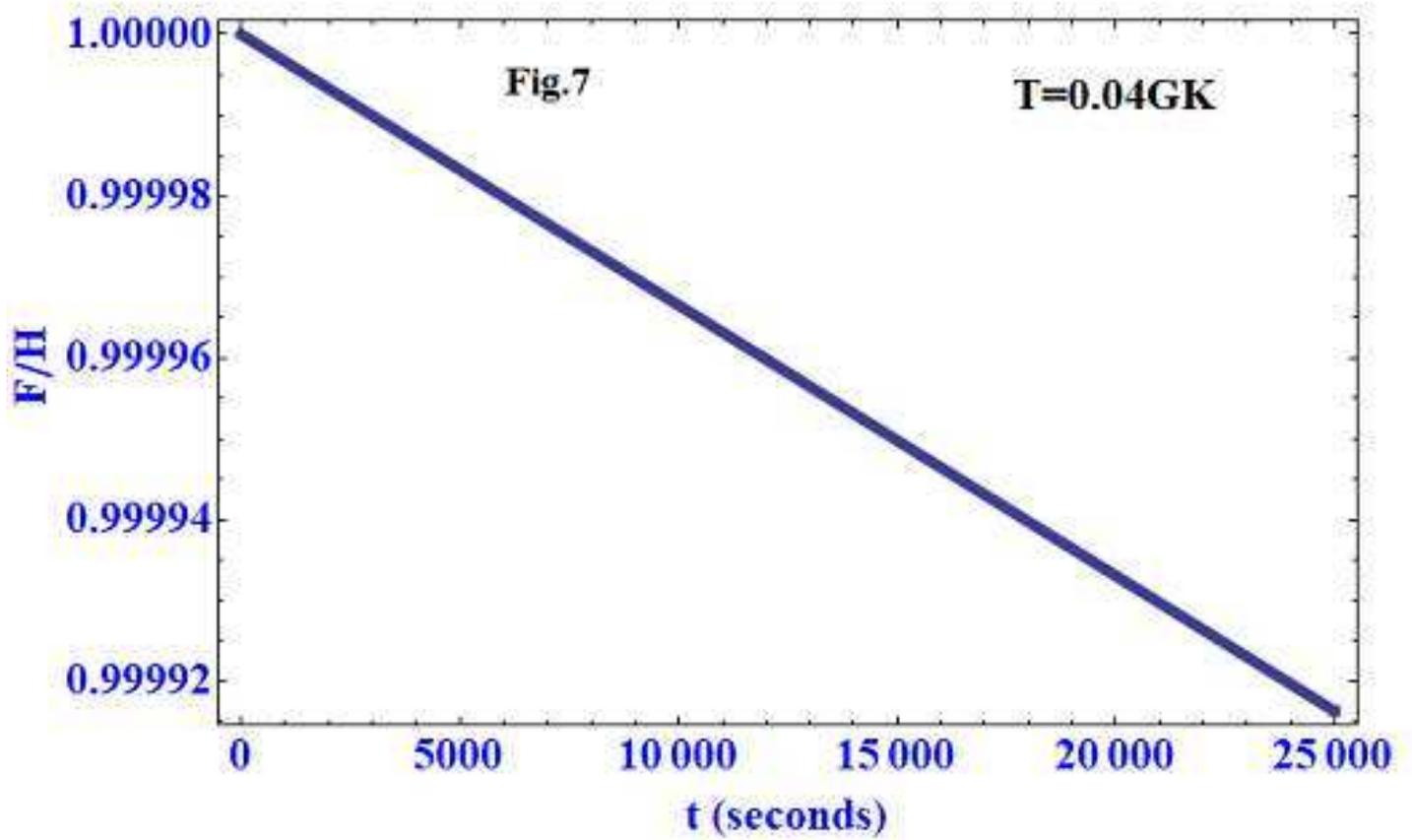
\end{document}